# Moonrise : Novel and Cartoon Writing System Built Upon Blockchain Systems

Hao Wang
CEO Office
Ratidar Technologies LLC
Beijing, China
Haow85@live.com

*Abstract*—Writing novels or drawing cartoons is a prolonged and interesting process that needs imagination and a lot of rethinking and rewriting. Blockchain systems have a very strong feature that tampering is not allowed for the system. In order to keep the revision history of novel writing / cartoon drawing, we apply blockchain systems such as HyperLedger to the problem and create a novel writer / cartoon illustrator system that is capable of keeping record of what has been written and greatly enhancing the writing performance of the author.

## I. INTRODUCTION

Blockchain system is one of the most notable inventions in recent years, including innovations in cryptocurrencies and ledger systems. The earliest invention of blockchains is BitCoin, whose price has skyrocketed to a level where the majority of global governments have kept a keen eye on the technology. Given the current investment hype on the technology, it is expected that without too long, we would be able to identify key points to improve the status quo.

The major bottleneck of blockchain system is the technical performance of the system. Throughput and latency are the 2 major evaluation metrics for the system. Both Bitcoin and ETH have had a long history of suffering from technical bottlenecks. Similarly, blockchain systems such as HyperLedger are also short of technical prominence when compared with conventional systems.

Blockchain systems are decentralized systems that are hacking proof and tamper proof, meaning that only when a majority of distributed nodes were hacked the system would break down, and also, the system could not be modified. The tamper proof property is highly useful because many applications in real world require record keeping. For example, SVN system for software writing, etc. , requires record keeping.

Novel writing has been surging in nowaday China, and cartoon creating likewise both in China and elsewhere in the world. With the help of internet platform for self-made novel writers, Chinese internet writers are able to produce a voluminous amount of literature works ranging from fantasies to martial art novels. Famous cartoon creators are distributed across both Europe, HKSAR (China), Japan and America, which are still in spotlight even in the information age of today.

Novel writing / cartoon drawing requires a lot of imagination work and rewriting, without the help of a record keeping system, it is hard to imagine that we could produce interesting novels. In this respect, blockchain system serves as an ideal choice for the work.

Entertainment company could create a enterprise-level HyperLedger system that keeps record of each version of revision of the novel writing / cartoon drawing work. This is especially helpful, not only for individuals, but also for the company as a whole.

In this paper, we propose to use HyperLedger Fabric to create a novel writing / cartoon drawing system that keeps record of each revision of the novel writing / cartoon drawing work. Our blockchain system is the first of its kind in the field of information systems.

## II. RELATED WORK

Blockchain is not a new topic. As old and glorious as any other field of computer science, blockchain has attracted a great number of international researchers. The earliest form of blockchain system is BitCoin [1]. Although the performance by technical metrics such as throughput and latency are far behind conventional systems, BitCoin remains one of the most successful cryptocurrencies and ledger systems.

Ethereum (ETH) [2] is the next wave of blockchain revolution, with millions of investors flooding in the cryptocurrency market purchasing the coin. It is actually one of the most lucrative businesses in the second decade of the 21 century. Dapps built upon the system have witnessed a surge during its development more than 5 years ago, but the eco-system of a full-fledged ETH community is still in shortage of investment of both money and technologies.

Decred [3] was haled as the 3rd generation cryptocurrency because of many of its superior properties. In addition, Decred does not support ICO, and therefore has much better image among the public. However, the cryptocurrrency is not as notable or successful as its 2 predecessors - namely BitCoin and Ethereum. Although the cryptocurrency is still in usage, the trade volume of the coin lags far behind other better known cryptocurrencies.

The key concept of blockchain technology is the consensus protocol algorithms. Famous consensus protocol algorithms include PBFT [4], Algorand [5], PoW [6], and PoS [7]. The basic ideas behind consensus protocol algorithms is to use distributed computing theories to create a mechanism that ensures safety and real-timeliness of the blockchain system.

There remain multitudes of issues related to blockchain systems besides the consensus protocol algorithms, such as Zero Knowledge Proof systems [8] - its theories and application remain a major research and development topic in the industry.

### III. SYSTEM DESIGN

We use HyperLedger Fabric to build a novel writing system, with codename Moonrise. The reason behind our choice of the technology is the following : 1. HyperLedger is not a public chain, which means that with a handful of distributed nodes, we are able to build a reliable system for novel writing workers. This is especially useful for entertainment firms. There are many cartoon and anime firms in Japan and America, and Europe likewise. Our Moonrise project is not only helpful for the novel writing companies, but also for such entertainment companies as well.

The system workout of HyperLedger Fabric is very simple and straightforward : 1. The HyperLedger Fabric system uses only a couple of definitions of data definitions : Participant is the only single most important concept element ; 2. The programming of a HyperLedger Dapp is highly trivial - the definition of participants is the core programming asset in the system; 3: Finishing up a whole Dapp system requires simple programming supported by multiple kind of languages.

We define the participants of the system in the following ways : 1. Novel / cartoon ID : The ID values of novels or cartoons ; 2. Revision number : The revision number values of novels or cartoons ; 3. Novel /cartoon content hash : The content has of novels or cartoons.

The implementation of the system design is as simple as the participant definition. Without much focus on the implementation details of the Moonrise system, we are able to generate a robust and holistic blockchain system that are not only functional but also tamper-proof.

### IV. WHY HYPERLEDGER FABRIC ?

HyperLedger Fabric is not a public chain. This is one of the utmost reasons why we take full advantage of HyperLedger Fabric to enhance the user experiences of author writing. As a permissioned blockchain, HyperLedger Fabric can be deployed over a multitude of distributed nodes with-in the company, and plus, without 2/3 of the nodes being hacked, the system would not be brought down by any intruders if Byzantine protocol is the used consensus protocol algorithm.

HyperLedger Fabric is faster than Ethereum by a magnitude of scale. In order to catch up with the conventional data management system such as MySQL or NoSQL databases, HyperLedger Fabric is one of the most prominent technologies that is expected to be able to catch up with conventional data technologies in speed.

The core principle behind the HyperLedger Fabric is that the execution workflow of the technology is different from other blockchain system such as Ethereum. Unlike the order-execute-validate paradigm deployed in other blockchain systems, HyperLedger Fabric uses the order-execute-validate paradigm, which is one of the major reasons why HyperLedger Fabric is much faster than other blockchain technologies such as BitCoin and Ethereum.

### V. NEW IDEA V.S OLD PARADIGMS

HyperLedger based novel writing / cartoon creating systems is a new idea in the field of blockchain systems. This conclusion is a bit shocking given that so much investment has been spent on the research and development of the technology. Most investors of the technology still focus only on profiteering, rather than a much healthier application of the technology.

The difficulty of developing a stronger blockchain ecosystem is we need to divert people from the cryptocurrency aspect of the technology to the Dapp application development. Cryptocurrency, especially that of public blockchain, is just a financial derivative that needs regulation. However, the technological property of the blockchain system leads to innovation in information security, hardware design, data management system development, etc. The advancement of the technology could even spill over to other sectors of the industry or economy. We should never underestimate the importance and impact of the blockchain system, even though much fewer scientists have focused on developing the technology for Dapps and other usage rather than financial purposes.

One notable old technology related to blockchain is Riak, which is a tamper proof database system whose basic principle is similar to blockchain. We would suggest scientists and engineers to refocus on old technologies such as Riak for better understanding to create a tamper proof technology that could even defeat the technology of blockchains. Another notable technology is Infobright. As a log system, data in Infobright is also tamper proof. Unlike blockchain system, Infobright could deal with large volume of datasets, and is also much faster than blockchain systems.

Our analysis of the blockchain market and technology is that currently blockchain community focuses on utilizing improvements over consensus protocols to lift the speed of the technology. Knowledge about other information systems such as Riak or Infobright have been largely ignored in many aspects of the system design. An entirely new paradigm of the blockchain system that either eliminates the old consensus protocol design, or merges the consensus protocol design with other technologies should be better studied for future generation of the blockchain technology.

Financial services could not support a healthy blockchain community because unexpected government regulation could lead to crush down of the eco-system at any time with an unbearable consequences for the investors. Government needs to divert the focus of the financial functionality of the technology to the software engineering part of the eco-system, as we have emphasized over and over again in this manuscript.

## VI. BLOCKCHAIN - WHO SHOULD INVEST AND WHO SHOULD BUY

Investment hype on blockchain has been replaced by investment in other sectors such as drug discovery. The investment on the eco-system has largely been driven by private sectors, although economic boost plans of governments sometimes cut in to improve the development of the research and technology of the blockchain.

To divert the investment from the financial functionality to the technical functionality, new investors should emerge and take the leading role. Contrary to the old investors, new investors should focus more on making huge improvement over the technicality soundness of the technology. The investment should be both huge and long-run. Without technical improvement, blockchain technology will degenerate into a financial derivative tool, which is not of much interest to the technical community and the society in a whole.

Who should pay for the blockchain technologies ? Tamper-proof products have lots of customers, especially in the financial institutes, critical social entities, etc. Customers need to have strong belief in the future of the blockchain company in order to make purchase. Positive propaganda of the technology is a must if more people are expected to purchase the technology. The current negative image of the blockchain technology as a financial profiteering tool greatly damages the market and customers' belief. In order to make up for the loss of the public image, it might take many years before the public recover from the negative sentiments and impressions.

A noted example of purchase of blockchain technologies is the Shandong (China's local province) public hospitals' purchase of blockchain empowered health insurance plans. Government empowered large scale purchases could lead to better improvement of the eco-system, and enforce the positive image of the technology to the public. Private buyers of the blockchain technology are still rarely seen in the industry, as blockchain is a distributed technology that requires more than 1 user.

## VII. CONCLUSIONS

In this paper, we propose a new blockchain system that enables efficient and effective novel writing / cartoon creating. Unlike conventional document / image processing systems, such as Microsoft Word package or image processing / animation creating systems, blockchain powered system are both secure and tamper-proof. The security feature not only protects the authors from malicious hackers, but also makes multiple distributed copies of the data for safe-keeping. The tamper-proof property guarantees that each revision of the literature is preserved for future reference, which is of great necessity for novel writing / cartoon creating.

In future work, we would like to explore the possibilities of creating other Dapps with stronger and better features for applications ranging from author writing / cartoon creating / movie making to other arena including financial assistance facilities to legal arbitration products.

Blockchain is the future of many industries unless the negative image of the public around cryptocurrencies is eliminated. We should never expect a consumer technology to be widely adopted or purchased unless the public have strong belief in the future of the technology. Blockchain is no exception.

### Acknowledgement

Human civilization is in a locked-up state unless wealth and opportunities are distributed more evenly. This is a conclusion discovered by the author in the year 2021, with nearly 10 publications followed suit. Laughably, these research papers might not elicit enough public awareness because they were published at international conferences / transactions whose prestigious levels are below SCI indexing. However, the author himself waves his hands to the opinion and criticism on the topic - I do respect the evaluation of the academic metrics on academic performance, but the academic circle do make their own judgement not only based on whether you are SCI or not, especially within the American academic circle.

I was admitted to a Harvard Online Business Program several years ago, but with objections from some sources, I didn't take the course. Harvard is the dream school of nearly every student in the world, but other considerations a l s o weighed in when I evaluated my destination. I did discuss the possibility of whether I should take the course or not, but I listened to others' opinion that online program is not as important or informative as on-site education, so I abandoned.

I was granted international research awards 4 times since 2008, and I am on my way to acquire money investment for my own startup. This year, I took CATTI Cup English Translation Contest in China. Just as last year, more than 80K contestants took part in the competition. In the preliminary round, I got two 76 for both written and oral contests, and the passing line for the written contest is 76. I was very surprised about the three 76 scores, so I didn't regret when I missed the next round because I was occupied with other work and I forgot to take the test.

Deep Learning is near its end in 2023 with large models consuming so many resources that only very few hands are capable of R&D for deep learning. However, the matrix operations of deep learning technologies could be reduced to simple matrix shallow models if you combine the tens of hundreds of matrix multiplications into one matrix and reduce the complexity of the model.

Once in a while, I double checked my book list and realized that deep insight should be the next hype in our society, rather than deep learning models. Deep insight means that we take more responsibility in publishing, business model building and policy making. We need deeper knowledge if our society wants leap-forwarding progress within the next decades.